\documentclass[aps,prb,superscriptaddress,showpacs,twocolumn,floatfix]{revtex4}
\usepackage{graphicx}
\usepackage{dcolumn}

\begin{document}
% You should use BibTeX and apsrev.bst for references
\bibliographystyle{apsrev}

%\preprint{}

\title{High Magnetic Field NMR Studies of LiVGe$_2$O$_6$, a quasi
1-D Spin $S = 1$ System.}
% Optional argument for running titles on pages
%\title[]{}


% other informatio
%\author{}
%\thanks{}
%\affiliation{}

\author{P.~Vonlanthen}
\affiliation{Department of Physics and Astronomy, University of
California at Los Angeles, Los Angeles, CA 90095-1547, U.S.A.}
\author{K.~B.~Tanaka}
\affiliation{Department of Physics and Astronomy, University of
California at Los Angeles, Los Angeles, CA 90095-1547, U.S.A.}
\author{Atsushi~Goto}
\affiliation{Department of Physics and Astronomy, University of
California at Los Angeles, Los Angeles, CA 90095-1547, U.S.A.}
\affiliation{National Institute for Materials Science, Tsukuba,
Ibaraki 305-0003 JAPAN }
\author{W.~G.~Clark}
\affiliation{Department of Physics and Astronomy, University of
California at Los Angeles, Los Angeles, CA 90095-1547, U.S.A.}
\author{P.~Millet}
\affiliation{Centre d'Elaboration des Mat\'{e}riaux et d'Etudes
Structurales, 29 rue J. Marvig, 31055 Toulouse Cedex, France}
\author {J.~Y. Henry}
\affiliation{Centre d'Etudes Nucl\'{e}aires, DRFMC/SPSMS/MDN,
F-38054 Grenoble Cedex 9, France}
\author{J.~L.~Gavilano}
\affiliation{Laboratorium f\"{u}r Festk\"{o}rperphysik,
ETH-H\"{o}nggerberg, CH-8093 Z\"{u}rich, Switzerland }
\author{H.~R.~Ott}
\affiliation{Laboratorium f\"{u}r Festk\"{o}rperphysik,
ETH-H\"{o}nggerberg, CH-8093 Z\"{u}rich, Switzerland }
\author{F.~Mila}
\affiliation{Institut de Physique Th\'{e}orique, Universit\'{e} de
Lausanne, 1015 Lausanne, Switzerland}
\author{C.~Berthier}
\affiliation{Grenoble High Magnetic Field Laboratory, BP 166,
38042 Grenoble Cedex 9, France}
\author{M.~Horvatic}
\affiliation{Grenoble High Magnetic Field Laboratory, BP 166,
38042 Grenoble Cedex 9, France}
\author{Yo~Tokunaga}
\affiliation{Grenoble High Magnetic Field Laboratory, BP 166,
38042 Grenoble Cedex 9, France}
\author{P.~Kuhns}
\affiliation{National High Magnetic Field Laboratory, Tallahassee,
Florida 32310 }
\author{A.~P.~Reyes}
\affiliation{National High Magnetic Field Laboratory, Tallahassee,
Florida 32310 }
\author{W.~G.~Moulton}
\affiliation{National High Magnetic Field Laboratory, Tallahassee,
Florida 32310 }


\date{\today}

\begin{abstract}

We report $^{7}$Li pulsed NMR measurements in polycrystalline and
single crystal samples of the quasi one-dimensional $S=1$
antiferromagnet LiVGe$_2$O$_6$, whose AF transition temperature is
$T_{\text{N}}\simeq 24.5$~K. The field ($B_0$) and temperature
($T$) ranges covered were 9-44.5~T and 1.7-300~K respectively. The
measurements included NMR spectra, the spin-lattice relaxation
rate ($T_1^{-1}$), and the spin-phase relaxation rate
($T_2^{-1}$), often as a function of the orientation of the field
relative to the crystal axes. The spectra indicate an AF magnetic
structure consistent with that obtained from neutron diffraction
measurements, but with the moments aligned parallel to the
$c$-axis. The spectra also provide the $T$-dependence of the AF
order parameter and show that the transition is either second
order or weakly first order. Both the spectra and the $T_1^{-1}$
data show that $B_0$ has at most a small effect on the alignment
of the AF moment. There is no spin-flop transition up to 44.5~T.
These features indicate a very large magnetic anisotropy energy in
LiVGe$_2$O$_6$ with orbital degrees of freedom playing an
important role. Below 8~K, $T_1^{-1}$ varies substantially with
the orientation of $B_0$ in the plane perpendicular to the
$c$-axis, suggesting a small energy gap for magnetic fluctuations
that is very anisotropic.

\end{abstract}
%\keywords{}

\maketitle

% body of paper here
\section{introduction}

Recently, a new quasi 1-D spin $S=1$ system, LiVGe$_2$O$_6$, has
been the object of intensive
experimental\cite{millet1999,gavilano2000,lumsden2000} and
theoretical investigations.\cite{mila2000,lou2000,liu2001} It has
an antiferromagnetic phase transition at about 25 K and the
expected Haldane gap is either absent or strongly suppressed.
Quantum chemistry calculations\cite{mila2000} indicate that a
second-order splitting $\Delta_{\text{CF}}$ of the $t_{\text{2g}}$
orbitals may play a dominant role in this system. Our new
measurements indicate that $\Delta_{\text{CF}}$ might be much
smaller than previously thought,\cite{millet1999} leading to a
large uniaxial magnetic anisotropy and orbital fluctuations.

It has been established by neutron diffraction measurements that
the low temperature phase has a rather simple, long-range
antiferromagnetic order.\cite{lumsden2000} In this paper, we
report the results of a number of different NMR measurements on
this material. We also address several important questions about
the phase transition which remained open previously, including the
order of the phase transition, the size and origin of the energy
gap in the magnetic excitation spectrum below the N\'{e}el
temperature, and the orientation of the magnetic moments in the
antiferromagnetic phase. Many of the results reported here were
obtained on powder samples. Some of the more recent measurements
were made on single crystal samples.

The $^7$Li NMR measurements we report include NMR spectra, the
spin-lattice relaxation rate $T_1^{-1}$, and the spin-spin
relaxation rate $T_2^{-1}$, at magnetic fields $B$ between 9.0 and
44.5~T and temperatures $T$ over the range 1.7 - 300~K. In spite
of various attempts to observe the resonance signal of $^{51}$V
nuclei above the transition and at the lowest temperatures in the
AF-phase, only a tiny spurious signal could be detected in the
polycrystalline sample and no signal at all was found in the
single crystal samples. The 9.0~T measurements were made at UCLA,
the measurements between 23 and 44.5~T were done at the NHMFL in
Tallahassee, and measurements at 12 T were performed at the GHMFL
in Grenoble. We have extended previous NMR measurements on a
polycrystalline powder sample\cite{gavilano2000} to much lower
temperatures as well as to much higher magnetic fields.
Furthermore, we present the first NMR measurements on
LiVGe$_2$O$_6$ single crystals as a function of the polar and
azimuthal angles, which give new insides on the low temperature
behavior of this system, where orbital degrees of freedom seem to
play an important role.

This paper is organized as follows. First, we describe the
preparation of the samples and the measurement procedures. Then,
we present the experimental results and a partial interpretation
of some of them. In the subsequent discussion we address issues
concerning the magnetic structure, the phase transition, the
relaxation rate and the influence of orbital degrees of freedom.

\section{Samples and experimental methods}

The LiVGe$_2$O$_6$ powder sample was prepared as described by
Millet \textit{et al.}\cite{millet1999} The single crystal samples
were synthesized at the Centre d'Etudes Nucl\'{e}aires in Grenoble
using a flux of GeO$_2$:Li$_2$B$_4$O$_7$ with the molar ratio 8:1.
After reducing the V$_2$O$_5$ with H$_2$ and using a slow increase
of the temperature up to 720~C, the compound LiVGe$_2$O$_6$ was
obtained with a thermal treatment of the components up to 800~C
under Ar with 2 \% vol H$_2$. Then, a mixture of 70~\%wt of the
flux and 30~\%wt of the compound was put in platinum crucible and
heated for 1 day at 970~C. After that, it was slowly cooled at the
rate of 2~C per hour to 780~C, after which the power to the
furnace was switched off and allowed to cool to room temperature.
Finally, the products were washed with boiling water. Pale green
needles were obtained, the maximum size of which was approximately
1~mm $\times$ 0.10~mm $\times$ 0.050~mm. The typical dimensions of
the samples used for our NMR measurements were $700~\mu$m $\times$
$100~\mu$m $\times$ $50~\mu$m, which corresponds to a mass of
about 15~$\mu$g and $\sim~3 \cdot 10^{16}$ $^7$Li spins. The small
NMR coils used for most of the single crystal work were a few
turns of 25~$\mu$m diameter insulated copper wire wound tightly
onto the sample.

LiVGe$_2$O$_6$ crystallizes in the monoclinic system, space group
$P2_1/c$.\cite{millet1999} The chains of VO$_6$ octahedra are
parallel to the $c$-direction and are connected to their neighbor
chains only by two GeO$_4$ tetrahedra. There is a very small
coupling perpendicular to the chains. The Vanadium atoms are
located in distorted oxygen octahedra and the three
$t_{\text{2g}}$ orbitals are split into a low-lying doublet
($d_{xy},d_{yz}$) and a single orbital ($d_{xz}$) at an energy
$\Delta_{\text{CF}}$ above the doublet.

All of the NMR results reported here were performed on the $^7$Li
nuclei using standard spin-echo techniques carried out with a
spectrometer and probes built at UCLA. The NMR spectra were
obtained by frequency-shifted and summed Fourier transform
processing\cite{clark1995} with fixed applied magnetic fields
between 9.0 and 44.5 Tesla.  Rotation of the field alignment about
one axis during the measurements was done by placing the sample
and NMR coil on a goniometer platform whose orientation was
controlled from the top of the probe. Further rotation about a
second axis perpendicular to the goniometer rotation axis was
carried out by changing the placement of the coil and sample on
the goniometer platform when the probe was out of the cryostat. We
estimate that the absolute accuracy of the corresponding angle
settings was approximately $\pm$10~deg and that the precision in
changing the angle with the goniometer was $\pm$0.5~deg. Part of
the uncertainty in the absolute angle was associated with the
small size of the samples and part of it from thermal contraction
in the goniometer control upon cooling from room temperature to
low temperatures.

An unsuccessful attempt was made to observe the $^{51}$V NMR
signal in a single crystal sample. A thorough search was done by
sweeping the resonance frequency in the absence of any external
field at 4.2~K as well as sweeping $B$ (aligned along the
$c$-axis) between 0 and 14~T with $T$ in the range 1.5~K-5~K at
the fixed frequencies 200, 300 and 550~MHz. We attribute the lack
of a signal, which otherwise should have been rather intense, to
values of $T_1$ or $T_2$ that were less than the approximately
2~$\mu$s dead time of the NMR spectrometer. It may be that
extending such measurements to much lower $T$ will reveal this
$^{51}$V signal.

The $^7$Li $T_1^{-1}$ measurements were performed by first rotating the
nuclear magnetization out of equilibrium by a short saturation
chain of rf-pulses, then waiting a variable recovery time, $t$,
and finally measuring the integrated spin-echo intensity, $m(t)$.
As discussed below, the quadrupolar splitting of the NMR line is
very small, about 15~kHz, so that a single exponential form is
expected for $m(t)$ as long as all parts of the sample have the
same value of $T_1^{-1}$. To monitor any deviation from the single
exponential behavior, we used a stretched exponential to fit our
data:
\begin{equation}\label{mt}
  m(t)= m_{\infty} + (m_0-m_{\infty})\exp(-t/T_1)^\beta,
\end{equation}
where $m_0$ and $m_{\infty}$ are the nuclear magnetization at
$t=0$ just after the saturation sequence and the equilibrium
magnetization, respectively. The fit parameter $T_1$ is the single
time that characterizes the recovery of the magnetization. It is
the time for the quantity $[m_{\infty}-m(t)]/[m_{\infty}-m_0]$ to
decay to 1/$e$. The exponent $\beta$ reflects the width of the
distribution of relaxation rates. For $\beta=1$, it corresponds to
a single exponential and as $\beta$ decreases from 1, it
represents a progressively broader distribution. The $T_1^{-1}$
measurements were done at Larmor frequencies between 149 and 762
MHz and applied magnetic fields between 9.0 and 41.5 T. For some
of the measurements at high magnetic fields below 3~K only the
beginning of the recovery curves were measured and the parameter
$m_{\infty}$ was set using values from the measurements between 10
and 3~K and the inverse temperature dependence of $m_{\infty}$.

Our $T_2^{-1}$ measurements were done at 148.981~MHz in a field of
9.0~T. The pulse sequence used was a $\pi/2$ preparation pulse
applied to $m_{\infty}$ followed a time $\tau$ later by a second
pulse whose angle was set to maximize the amplitude of the echo.
The integral of the spin echo signal was recorded as a function of
$\tau$. The decay of the signal was analyzed using the function:
\begin{equation}\label{t2}
  m(2\tau)=m(0)\exp-(2\tau/T_2)^\beta ,
\end{equation}
where $\beta$ is the stretched exponential parameter between 1
(exponential decay) and 2 (gaussian decay). For the powder sample
the $\beta$ parameter was usually left free during the fit, and
resulted in values around 1.4. The spin-echo amplitude for the
single crystal measurements was modulated by the quadrupolar
interaction, which caused strong deviations of $m(2\tau)$ from an
exponential decay. In this case, the value of $\beta$ used in the
analysis was fixed at 2.

\section{Experimental results}\label{expresults}

\subsection{NMR spectra of polycrystalline LiVGe$_2$O$_6$} \label{spec}

\begin{figure}[htbp]
\includegraphics[width=90mm]{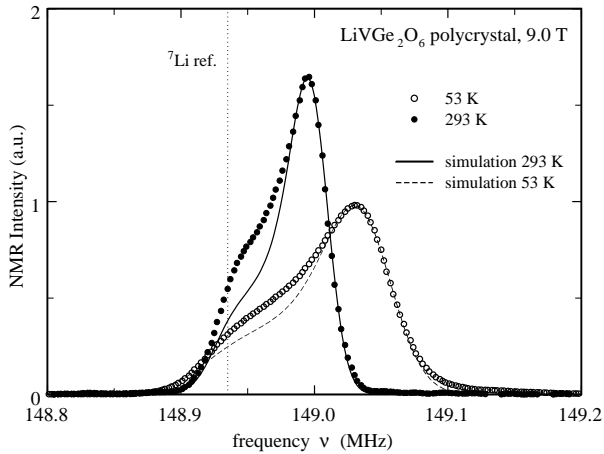}
\caption{\label{sphighT} $^7$Li NMR spectra in the paramagnetic
phase of polycrystalline LiVGe$_2$O$_6$. The dotted line is the
expected position of $^7$Li in a reference compound like LiCl. The
solid and dashed lines are simulations (see text).}
\end{figure}

Figure~\ref{sphighT} shows two NMR spectra of the powder sample in
the paramagnetic regime, i.e. at $T>T_{\text{N}}\approx
\text{25~K}$. The experimental points are indicated by the symbols
and the solid and dotted lines are fits to a model of a
polycrystalline powder in the presence of an axially symmetric,
anisotropic shift,\cite{carter1977} as discussed below.

For the simulation of the asymmetric NMR spectra in the
paramagnetic regime we assumed each V ion to have a moment along
the applied magnetic field, whose magnitude is independent of the
orientation. Hence, the anisotropy of the $g$-factor was not taken
into account. It was, however, verified that when
$1>g_{\perp}/g_{\parallel} \geq 0.5$, results similar to those for
an isotropic $g$-factor are obtained. The corresponding magnetic
field at the Li sites was then calculated by adding all the dipole
contributions of the V ions in a sphere of about 5~nm diameter
around the Li ion. It was verified that modifying the diameter of
the sphere does not have any effects on the results. The hyperfine
field at the Li sites cannot, however, be fully accounted for by
assuming a purely dipolar field of the V moments. An additional
isotropic hyperfine coupling of the order of 0.048
T/$\mu_{\text{B}}$ is needed. The latter may arise in a manner
similar to the superexchange interaction. By assuming a randomly
distributed powder, i.e., the direction of the applied magnetic
field is pointing along all possible directions of the unit
sphere, the Li spectra were then simulated at different
temperatures. From the simulations of the measured spectra, one
obtains for the average component $<M_{\text{Z}}>$ of the magnetic
moments along the direction of the applied field 0.048
$\mu_{\text{B}}$ at 300 K and 0.080 $\mu_{\text{B}}$ at 53 K,
respectively. Their ratio is what one expects from the temperature
dependence of the dc-susceptibility.\cite{gavilano2000}  The
simulations, shown in Fig.~\ref{sphighT} are based on only three
parameters: i) an isotropic hyperfine coupling, which is the same
for all the data, ii) the size of the magnetic moment on the V
ions, whose temperature dependence follows the dc-susceptibility
and iii) a gaussian broadening function. In view of the small
number of parameters the fits agree fairly well with the measured
data. Some deviations are observed in the low frequency part of
the signal; their origin is not yet understood.

\begin{figure}[htbp]
\includegraphics[width=85mm]{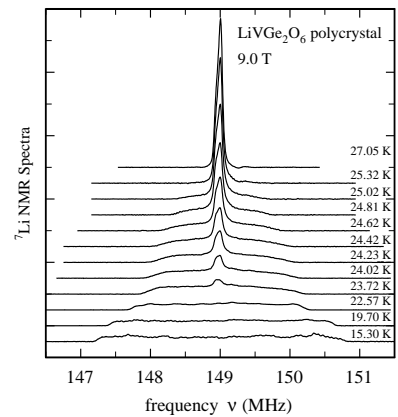}
\caption{\label{sprep} $^7$Li NMR spectra of polycrystalline
LiVGe$_2$O$_6$ at 9.0 T for $T$ between 15 and 27~K.}
\end{figure}

\begin{figure}[htbp]
\includegraphics[width=85mm]{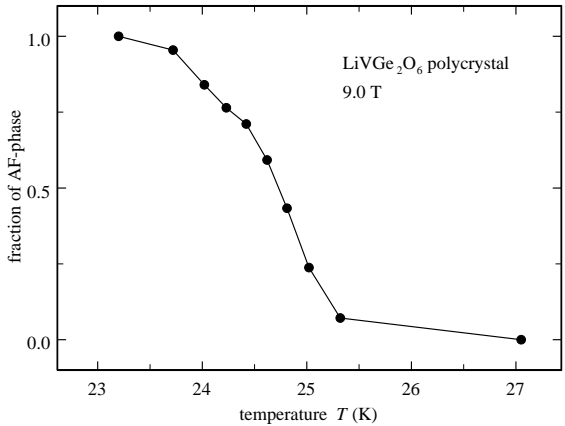}
\caption{\label{ratio} Fraction of the total NMR intensity in the
polycrystalline LiVGe$_2$O$_6$ spectra at 9.0 Tesla which is
attributed to the antiferromagnetic phase. }
\end{figure}

The 9.0 T NMR powder spectra near the transtion and in the ordered
phase are shown in Fig.~\ref{sprep}. They show a continuous
transfer of spectral weight from a narrow line in the paramagnetic
phase to a broad signal in the antiferromagnetic (AF) phase that
occurs over a narrow temperature range. As seen in
Fig.~\ref{ratio}, which shows the fraction of the intensity in the
AF phase, both phases coexist in a temperature range of about
1.5~K around 24.4~K.

The order parameter of the AF phase is the magnitude and
polarization of the AF moments associated with the V atoms. They
generate a corresponding magnetic field $\mathbf{B_{\text{i}}}$
 at the i-th $^7$Li sites, which can be calculated for a given AF
moment configuration. Depending on the configuration of the AF
state, $\mathbf{B_{\text{i}}}$  may have a sequence of values at
different $^7$Li sites or be given by a single value for all of
them. At the high magnetic fields used in our experiments, the
$^7$Li spins probe the static order parameter through the shift in
their spectrum, which is given by the component of
$\mathbf{B_{\text{i}}}$ that is parallel to the applied field
$\mathbf{B_{0}}$. For a randomly-oriented polycrystalline sample,
the NMR spectrum in the AF phase depends on the response of the AF
polarization to the varying orientation of the external field.

The NMR spectra in the AF phase below 23~K have a broad, nearly
rectangular shape. This shape is expected for a randomly oriented
powder spectrum if $\mathbf{B_{\text{i}}}$ has the same magnitude
at all $^7$Li nuclei, is parallel or antiparallel to a single
crystalline direction and maintains the same orientation with
respect to the crystalline axes for all orientations of
$\mathbf{B_{0}}$. For $B_{\text{i}}\ll B_{0}$, the field at the
nuclei is given by $B_{0} + B_{\text{i}}\cos(\theta)$, where
$\theta$ is the polar angle in spherical coordinates. Then, the
frequency shift is: $f=\gamma B_{\text{i}}\cos(\theta)$. The
probability ($dN$) that a particular value of $\theta$ occurs is:
$dN=1/2\sin(\theta)d\theta$ and the density of states in the
powder pattern is:
\begin{equation}\label{specsquare}
  \frac{dN}{df}=\left|\frac{dN}{d\theta}\right|\left|
  \frac{d\theta}{df}\right|=\frac{1}{2\gamma B_{\text{i}}}~,
\end{equation}
i.e., constant and results in a rectangular shape for the spectrum.

Since this shape is indeed observed, the powder spectra indicate
that the direction of the internal magnetization is not affected
by the orientation of the external magnetic field. From these
measurements (see Fig.~\ref{sprep}) we obtain the value
$B_{\text{i}}=0.106$~T.

\begin{figure}[htbp]
\includegraphics[width=85mm]{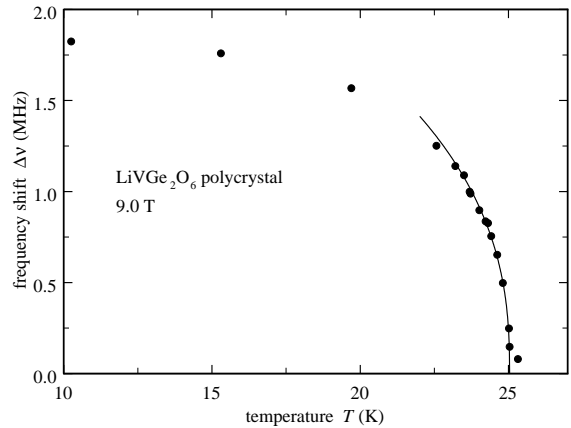}
\caption{\label{width} Frequency shift $\Delta\nu$ from the NMR
spectra of polycrystalline LiVGe$_2$O$_6$. The width in the
paramagnetic phase at 26~K has been subtracted. The solid line is
a fit to the data (see text).}
\end{figure}

In Fig.~\ref{width} the part of the NMR linewidth ($\Delta\nu(T)$)
proportional to the AF order parameter is plotted as a function of
temperature. The contribution of the width in the paramagnetic
phase is subtracted in quadrature from the AF contribution using
\begin{equation}\label{deltanuf}
  \Delta\nu=(\Delta\nu_{\text{AF}}^2-\Delta\nu_{\text{P}}^2)^{1/2}~,
\end{equation}
where $\Delta\nu_{\text{AF}}$ and $\Delta\nu_{\text{P}}$ are the
HWHM of the spectra in the AF phase and in the paramagetic phase
at 26~K, respectively. The solid line corresponds to a power law
behavior $\Delta\nu(T)\propto(T_{\text{N}}-T)^{0.4\pm 0.05 }$. The
onset of the broadening of the spectra occurs at 25.04~K, which we
identify as the N\'{e}el temperature $T_{\text{N}}$ for the
polycrystalline sample.

\subsection{NMR spectra of LiVGe$_2$O$_6$ single crystals}\label{specsingle}

\begin{figure}[htbp]
\includegraphics[width=85mm]{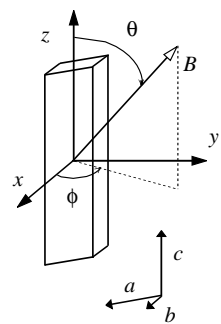}
\caption{\label{angle} Definition of various coordinates for a
typical LiVGe$_2$O$_6$ single crystal.}
\end{figure}

The coordinates shown in Fig.~\ref{angle} will be used to discuss
our measurements on the single crystal samples. They include the
crystalline axes $a$, $b$, and $c$, the cartesian axes $x$, $y$,
and $z$, and the spherical coordinates $\theta$~(polar angle) and
$\phi$ (azimuthal angle). X-ray measurements\cite{ravy2001} have
shown that the long dimension of LiVGe$_2$O$_6$ single crystals is
along the crystallographic $c$-direction. For all of our single
crystal NMR measurements, the value of $\phi$ is close to zero,
unless specified otherwise.

\begin{figure}[htbp]
\includegraphics[width=85mm]{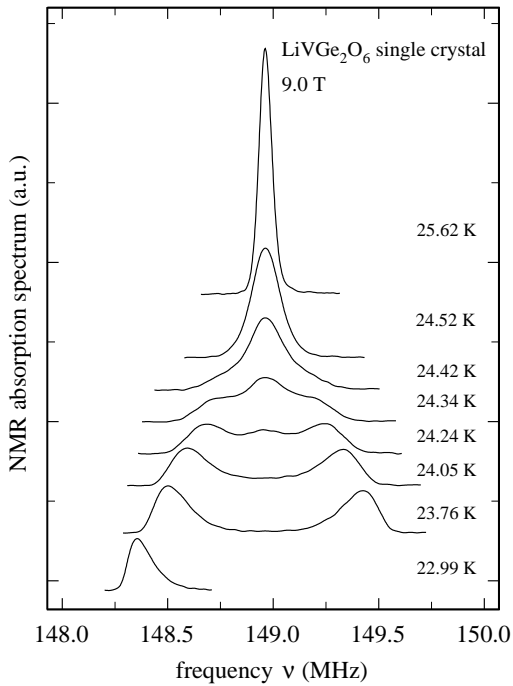}
\caption{\label{spec30} Single crystal $^7$Li NMR spectra of
LiVGe$_2$O$_6$ near $T_{\text{N}}$. The needle direction, which
corresponds to the crystallographic $c$-direction, is aligned
60$^{\circ}$ from the applied magnetic field.}
\end{figure}

Figure~\ref{spec30} shows NMR spectra of a LiVGe$_2$O$_6$ single
crystal at temperatures between 23 and 26~K for
$\theta\approx60^{\circ}$. The NMR spectrum is a single line in
the paramagnetic phase and two lines in the AF phase that
correspond to the two magnetically inequivalent $^7$Li sites. As
for the powder spectra, both phases coexist over a small range of
$T$. However, this range is substantially narrower with a value
$\simeq 0.3$~K for the single crystal.

As for the powder sample, the single crystal linewidths are rather
broad near $T_{\text{N}}$. Just above the transition,
$\Delta\nu(T)$ is about 36~kHz compared to 19~kHz at 38~K. Since
the dc-susceptibility increases with increasing $T$ in this range,
the opposite change in the linewidth may indicate that there are
slow, short range fluctuations above the phase transition that
enhance the linewidth somewhat. Furthermore, in the whole
considered temperature range, the linewidths of the single crystal
signals are consistent with the width of the broadening function
used to fit the spectra of the polycrystalline sample. The large
NMR line width seen just below $T_{\text{N}}$ may also be caused
by a distribution of $T_{\text{N}}$ in the sample.

\begin{figure}[htbp]
\includegraphics[width=85mm]{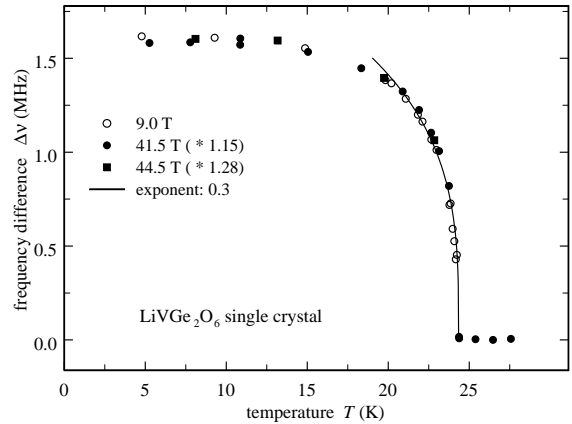}
\caption{\label{TLHorderpar} Frequency shift $\Delta\nu$ of the
$^7$Li peaks in the AF phase of LiVGe$_2$O$_6$ from the peak in
the paramagnetic phase as a function of $T$. For comparison, the
measurements at 41.5 and 44.5 Tesla have been multiplied by the
factors 1.15 and 1.28 respectively. The solid line is a fit to the
data (see text).}
\end{figure}

In Fig.~\ref{TLHorderpar}, $\Delta\nu(T)$, representing the
splitting of the two peaks in the $^7$Li NMR spectrum, is plotted
as function of $T$ for  $B_{0}=$ 9.0, 41.5 and 44.5 T, with
$\theta=$~30$^{\circ}$ at 9.0 T and approximately 20$^{\circ}$ for
the measurements at 41.5 and 44.5~T. The data obtained at 41.5 and
44.5~T have also been multiplied by 1.15 and 1.28, respectively,
to bring the curves together. This indicates a somewhat reduced
value of the order parameter for the measurements in higher
fields, an aspect that will be discussed later. For the same
reason, the values of $T$ have been reduced by 0.6~K and 1.6~K for
the measurements at 41.5~T and 44.5~T, respectively. We attribute
these adjustments in $T$ as being caused by differences in the
instrumentation, which resulted in an uncertainty of about 1~K in
$T_{\text{N}}$ for the measurements at these higher values of
$B_{0}$.

Because of the large values of $B_{0}$, a substantial reduction of
$T_{\text{N}}$ proportional to $B_{0}^2$ is
expected.\cite{shapira1970} However, as can be seen in
Fig.~\ref{TLHorderpar}, no such reduction is observed. This seems
rather surprising because $g\mu_{\text{B}}B/k_{\text{B}}= 54$~K,
for $B=44.5$~T with $g=1.79$, i.e., about twice $T_{\text{N}}$. It
may be, however, that the value of $g=1.79$, obtained from the
paramagnetic susceptibility\cite{millet1999} is too large because,
as will be discussed later, the measurements of Lumsden \textit{et
al.},\cite{lumsden2000} as well as our own measurements, show that
the value of the ordered moment is only about
1.14~$\mu_{\text{B}}$. Furthermore, it is evident that a large
change in $B_{0}$ has no significant effect on the $T$ dependence
of the order parameter in the ordered phase, as the three curves
coincide. The solid line is a power law fit to the data just below
the transition with $\Delta\nu(T)\propto(T_{\text{N}}-T)^{0.3\pm
0.05}$. The value of the exponent is somewhat smaller than that
0.4 obtained for the polycrystalline sample. Qualitatively, such a
reduction is consistent with the smaller distribution of
$T_{\text{N}}$ inferred from the smaller temperature range over
which a coexistance of two phases is indicated in the single
crystal samples.

\subsection{$T_1^{-1}$ ($T$, $B$) for polycrystalline
 LiVGe$_2$O$_6$}\label{t1tb}

In this section we describe our spin-lattice relaxation rate
measurements at 9.0 and 23 Tesla performed on the polycrystalline
sample. Below $T_{\text{N}}$ the measurements were done at the
same frequency and magnetic field as above the transition. But
because of the very broad linewidth of about 3.5 MHz in the
AF-phase, only a small part of the NMR spectrum was covered by the
RF-pulses. The nuclei near resonance were those with $\theta
\approx90^{\circ}$; i.e., $\mathbf{B_{0}}$ close to lying in the
azimuthal plane.  The angle $\phi$ is distributed randomly over
$2\pi$.

\begin{figure}[htbp]
\includegraphics[width=85mm]{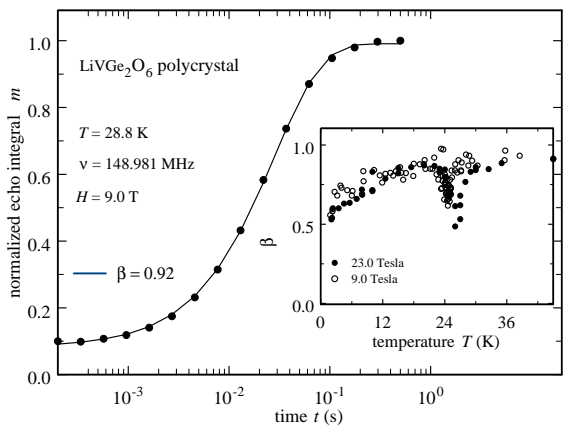}
\caption{\label{t1503l} Recovery of the magnetization during a
$T_{1}^{-1}$ measurement in polycrystalline LiVGe$_2$O$_6$. The
solid line is the fit of Eq.~\ref{mt} for $\beta = 0.92$. Inset:
Plot of $\beta$ as function of $T$ at 9.0 and 23~T.}
\end{figure}

Figure~\ref{t1503l} shows a typical magnetization recovery curve
(filled circles). The best fit using Eq.~\ref{mt} (solid line),
which yields $\beta=0.92$, is an excellent fit to the data. The
inset of Fig.~\ref{t1503l} shows that the $T$-dependence of
$\beta$ is essentially the same at both 9.0~T and 23.0~T. Except
for $T$ close to $T_{\text{N}}$, above 12~K the values obtained
for $\beta$ are close to 1 and correspond to a relatively narrow
distribution of $T_1^{-1}$. Near the transition and below 12~K,
$\beta$ considerably deviates from 1, which indicates a
substantially broader distribution of $T_1^{-1}$ at these
temperatures. We attribute the broadening of the distribution of
$T_1^{-1}$ close to $T_{\text{N}}$ to the distribution of
$T_{\text{N}}$ for the different parts of the sample mentioned
earlier. As discussed below for the single crystal measurements,
the large deviation from $\beta=1$ seen below 12~K reflects a
large, unexpected dependence of $T_1^{-1}$ upon $\phi$.

\begin{figure}[htbp]
\includegraphics[width=85mm]{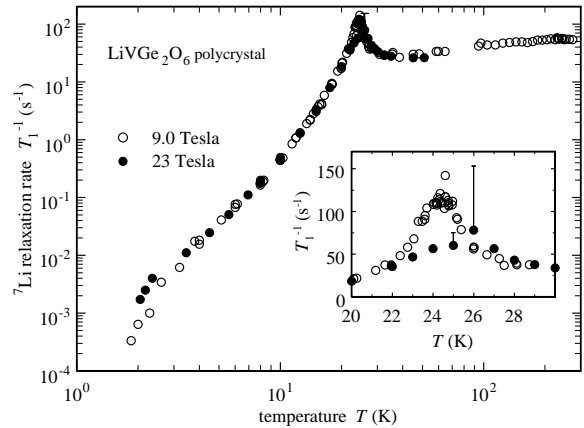}
\caption{\label{LiT1linlin} $^7$Li $T_1^{-1}$  as function of $T$
in polycrystalline LiVGe$_2$O$_6$ at 9.0 (open circles) and 23
Tesla (filled circles).}
\end{figure}

The $T_1^{-1}$ values obtained for 9.0~T and 23.0~T are shown in
Fig.~\ref{LiT1linlin}. Where $\beta$ is substantially less than 1,
a wide distribution of values for $T_1^{-1}$ is present.
Therefore, the plotted value is the one that represents the single
recovery rate that characterizes this distribution. Nevertheless,
from its $T$-dependence useful information on the dynamics of our
system can be obtained, even at the lowest values of $T$. For the
9.0~T measurements, well above $T_{\text{N}}$, $T_1^{-1}$ depends
only weakly on $T$, slowly increasing by about a factor 2 between
40 and 200~K and remaining almost constant between 200 and 300~K.
Therefore, as previously reported\cite{millet1999,gavilano2000},
there is no indication of a Haldane gap in this quasi 1-D system.
Below 40~K, $T_1^{-1}$ increases rapidly to a maximum value of
about 140~s$^{-1}$ at $T_{\text{N}}$, presumably due to critical
fluctuations near the transition.

Below 23~K, $T_1^{-1}$ drops very rapidly by about 6 decades to a
value near $3\cdot 10^{-4}$~s$^{-1}$ at 1.8~K. In the low $T$
regime, the effect of $B_{0}$ on $T_1^{-1}(T)$ is very weak.
Because the value of $g\mu_{\text{B}}B/k_{\text{B}}$ for 23~T is
close to $T_{\text{N}}$ this result was unexpected, as was the
small influence of $B_{0}$ on the spectra shown in
Sec.~\ref{specsingle} which are also affected only weakly by
$B_{0}$.

\subsection{$T_1^{-1}(T)$ of a LiVGe$_2$O$_6$ single
crystal}\label{t1single}

\begin{figure}[htbp]
\includegraphics[width=85mm]{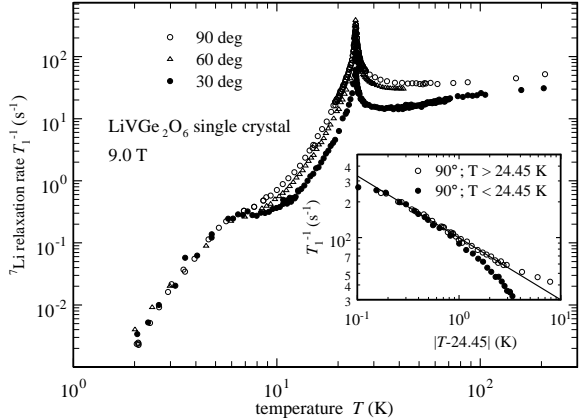}
\caption{\label{t1or} $^7$Li $T_1^{-1}$ as function of $T$ in a
single crystal of LiVGe$_2$O$_6$. The three values of the polar
angle $\theta$ are 90$^{\circ}$ (open circles), 60$^{\circ}$ (open
triangles), and 30$^{\circ}$ (filled circles). Inset: $T_1^{-1}$
as function of $|T-24.45|$ above (open circles) and below (filled
circles) $T_{\text{N}}$ for $\theta \approx 90^{\circ}$. }
\end{figure}

Figure~\ref{t1or} shows measurements of $T_1^{-1}$ as a function
of $T$ for a single crystal of LiVGe$_2$O$_6$ for
$\theta=90^{\circ}$, 60$^{\circ}$, and 30$^{\circ}$, with
$\phi\sim0^{\circ}$. Over the whole temperature range, no
significant deviations from $\beta=1$ were observed. This behavior
indicates that unlike the polycrystalline sample, there is no
distribution in $T_1^{-1}$; i.e., the relaxation follows a single
exponential. In principle, this result should make a detailed
interpretation of the results more straightforward.

The general behavior of $T_1^{-1}(T)$ is similar to that of the
polycrystalline sample. From 200~K to 40~K the relaxation rate
slightly decreases by about a factor of two, but subsequently
increases by more than one order of magnitude towards
$T_{\text{N}}$, and decreases rapidly below $T_{\text{N}}$. The
inset of Fig.~\ref{t1or} shows a log-log plot of $T_1^{-1}$ for
$\theta \approx 90^{\circ}$ close to $T_{\text{N}}$ as a function
of $|T-T_{\text{N}}|$ with $T_{\text{N}}=24.45$~K. Close to the
transition, the data both below and above $T_{\text{N}}$, fall on
the same curve, given by
\begin{equation}\label{t1crit}
  T_1^{-1}\propto |T-24.45|^{-0.55}~.
\end{equation}
Although the exponent $-0.55$ is expected to reflect the critical
behavior of the AF transition in this material, it should be
interpreted with caution. Because the width of the transition
shown by the coexistence of both phases (Fig.~\ref{spec30}) over a
range of 0.3~K may indicate a distribution of $T_{\text{N}}$ in
the sample, it may be that the exponent is really a lower limit on
the rate of the divergence on approaching $T_{\text{N}}$.

\begin{figure}[htbp]
\includegraphics[width=85mm]{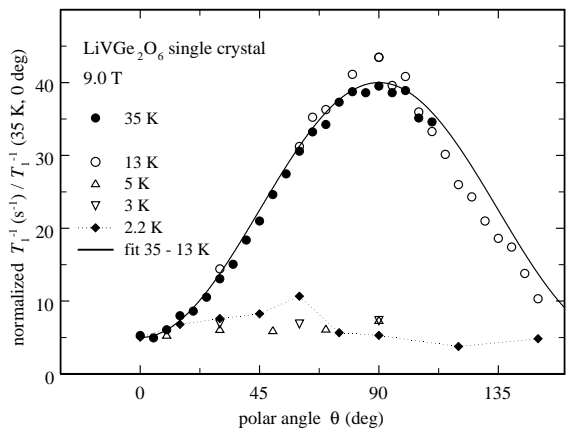}
\caption{\label{t1angle} Angular dependence of $T_1^{-1}$ in a
LiVGe$_2$O$_6$ single crystal at five temperatures between 35 and
2.2~K. The solid line is a fit to the data at 35 and 13~K (see
text).}
\end{figure}

Figure~\ref{t1or} also shows that the $\theta$-dependence of
$T_1^{-1}(T)$ on $T$ has a crossover from $\theta$ being
approximately independent of $\theta$ below 8~K to a strong
dependence above 10~K. This behavior is shown in more detail in
Fig.~\ref{t1angle}, where $T_1^{-1}$ is plotted as function of
$\theta$ for several values of $T$ both above and below 10~K.
Above 10~K, $T_1^{-1}$ is well described by
\begin{equation}\label{sin2}
  T_1^{-1}(\theta)= A(T) [ 35\sin^2(\theta)+5]~,
\end{equation}
where $A$ determines the magnitude of $T_1^{-1}$. The measurements
at 5 and 3~K do not show any dependence on $\theta$. At 2.2~K, the
moderate $\theta$-dependence of $T_1^{-1}$ is not well enough
established to draw useful conclusions.

\begin{figure}[htbp]
\includegraphics[width=85mm]{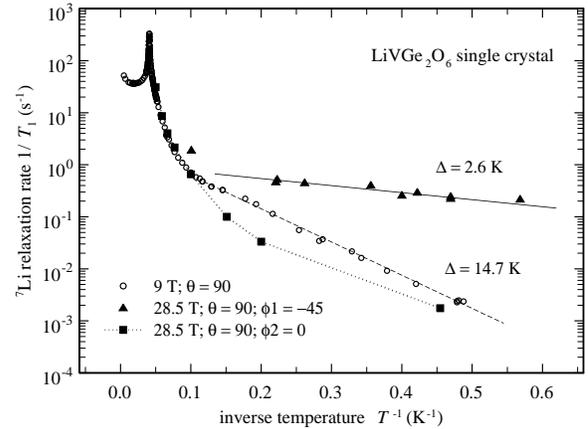}
\caption{\label{t1orient3} $^7$Li $T_1^{-1}$ as function of 1/$T$
in a LiVGe$_2$O$_6$ single crystal at 28.5~T and the polar angle
$\theta=90^{\circ}$ for two azimuthal angles $\phi_1=-45^{\circ}$
(filled circles) and $\phi_2=0^{\circ}$ (filled squares) and at
9.0~T at $\theta=90^{\circ}$ (open circles). The solid and dashed
lines indicate energy gaps of 2.6 and 14.7~K respectively. The
dotted line is a guide to the eye.}
\end{figure}

\begin{figure}[htbp]
\includegraphics[width=85mm]{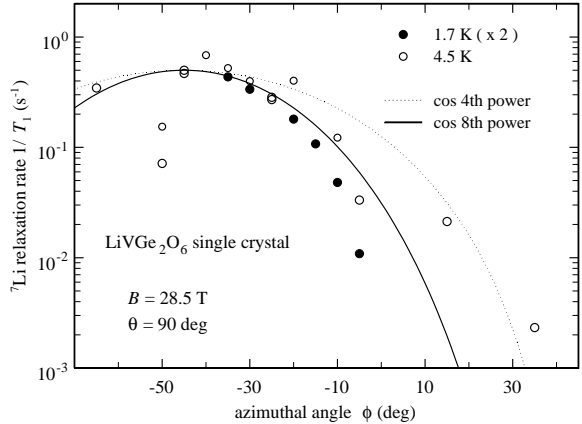}
\caption{\label{t1orient4} $^7$Li $T_1^{-1}$ as function of $\phi$
in a LiVGe$_2$O$_6$ single crystal at 28.5~T and
$\theta=90^{\circ}$ for $T=4.5$~K (open circles) and $T=1.7$~K
(filled circles).}
\end{figure}

Now we turn to measurements in which $\phi$ was varied and
$\theta$ is held fixed at $90^{\circ}$; i.e., $\mathbf{B_{0}}$ was
rotated in the azimuthal plane. For these measurements, the
direction corresponding to $\phi=0$ is always the same, but the
location of $\phi=0$ in the azimuthal plane is not known.
Figure~\ref{t1orient3} shows $T_1^{-1}$ at $\theta=90^{\circ}$ as
a function of 1/$T$ for $\phi_1=-45^{\circ}$ and
$\phi_2=0^{\circ}$ at 28.5~T and, for comparison at 9.0~T and
$\phi=0^{\circ}$. In addition the $\phi$ dependence of $T_1^{-1}$
for 4.5 and 1.7~K at $\theta=90^{\circ}$ and 28.5~T is shown in
Fig.~\ref{t1orient4}. A huge change of two decades in $T_1^{-1}$
is seen for a variation of only 30$^\circ$ in $\phi$.

These variations of $T_1^{-1}$ with $\theta$ and $\phi$ help to
explain the observation that $\beta \leq 1$ throughout the entire
range of $T$ for the polycrystalline sample (Fig.~\ref{t1503l}).
Below $\sim 8$~K, the variation of $T_1^{-1}$ as a function of
$\phi$ causes a very broad distribution of $T_1^{-1}$ in the
polycrystalline sample that is qualitatively consistent with the
established small values of $\beta$. Similarly, above
$T_{\text{N}}$, the narrower distribution of $T_1^{-1}$ caused by
its variation with $\theta$ (Fig. \ref{t1or}) results in a value
of $\beta$ that is slightly less than one. In the range 10-20~K,
the moderate range of $\theta$ near $90^{\circ}$ selected by the
rf pulse at the center of the spectrum and the variation of
$T_{\text{N}}$ in the sample are probably the major conditions
responsible for the measured value of $\beta$.

\begin{figure}[htbp]
\includegraphics[width=85mm]{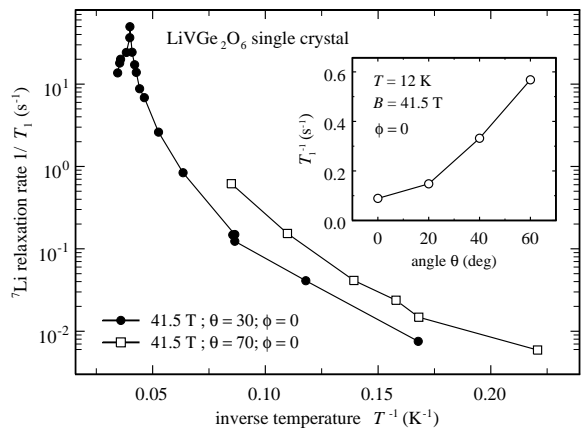}
\caption{\label{t1orient2} $^7$Li $T_1^{-1}$ as function of 1/$T$
in a LiVGe$_2$O$_6$ single crystal at 41.5~T for
$\theta=30^{\circ}$ (solid circles) and $\theta=70^{\circ}$ (open
squares. Inset: $\theta$-dependence of $T_1^{-1}$ at 12~K and
41.5~T.}
\end{figure}

Our $T_1^{-1}$ measurements at the highest $B_{0}$ of 41.5~T for a
limited number of angles and temperatures are shown in
Fig.~\ref{t1orient2} and its inset. No significant deviation from
the results at lower $B_{0}$ is evident. This behavior shows that
although the dynamical properties of the electron moments that
form the AF state are quite sensitive to the alignment of
$\mathbf{B_{0}}$, they are nearly independent of its magnitude up
to 41.5~T.

\subsection{$T_2^{-1}(T)$ of LiVGe$_2$O$_6$}\label{T2}

\begin{figure}[htbp]
\includegraphics[width=85mm]{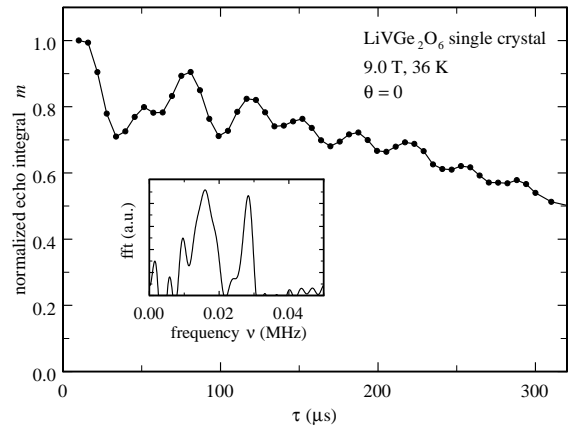}
\caption{\label{t2oscil} Spin-echo decay as a function of $\tau$
in a LiVGe$_2$O$_6$ single crystal at 9.0 Tesla. Inset: FFT of the
spin-echo decay.}
\end{figure}

Figure~\ref{t2oscil} shows the oscillatory behavior of the
amplitude of the spin-echo decay as a function of the pulse
spacing $\tau$ at 9.0~T, 36~K, and $\theta=0$ in a single crystal
sample. We attribute the modulation of the echo height to the
static quadrupole interaction of the $^7$Li nuclei. From the
period of the modulation, $\tau_m$, the quadrupole frequency,
$\nu_Q$, is obtained:\cite{abe1966} $\nu_Q=1/\tau_m$. Since the
orientation of the electric field gradient (EFG) tensor at these
sites is not known, our limited measurements do not provide an
exact value of $\nu_Q$. However, from the fourier transform of the
decay curve for a $T_2$ measurement with one of the shortest
periods (see inset Fig.\ref{t2oscil}), an approximate value of
$\nu_Q \approx$~15~kHz is obtained for the quadrupole frequency.
Actually, two frequencies $\nu_Q$ and $2\nu_Q$ are seen, as
expected for a nuclear spin $I=3/2$ system.\cite{abe1966} Also no
change in the modulation was observed over the entire range of $T$
that was covered. This behavior indicates that the EFG is constant
and that no structural change occurs at $T_{\text{N}}$, in
agreement with neutron and x-ray diffraction
measurements.\cite{lumsden2000}

\begin{figure}[htbp]
\includegraphics[width=85mm]{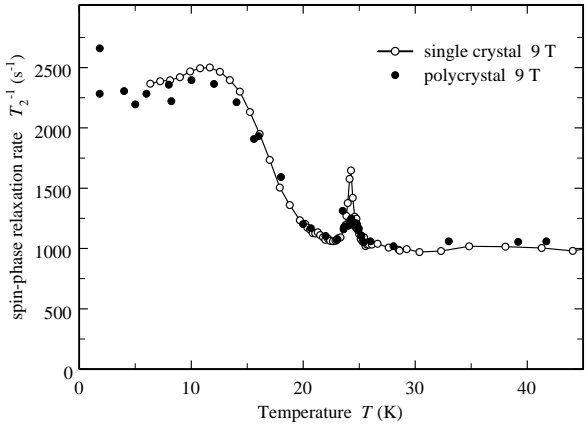}
\caption{\label{LiT2} $^{7}$Li $T_2^{-1}$ as function of $T$ at
9.0~T in polycrystalline LiVGe$_2$O$_6$ (filled circles) and a
LiVGe$_2$O$_6$ single crystal (open circles).}
\end{figure}

In Fig.~\ref{LiT2}, $T_2^{-1}$ as function $T$ is plotted at
$B_0=9.0$~T for both a polycrystalline powder sample and a single
crystal sample. Several features are seen in these measurements.
Except for very close to the transition,
$T_2^{-1}\simeq1000~$s$^{-1}$ above $T_{\text{N}}$. Also, there is
a narrow peak in $T_2^{-1}$ within 1~K of $T_{\text{N}}$ in which
$T_2^{-1}$ is enhanced by about 25 \% for the powder sample and by
70~\% for the single crystal one. We attribute these increases to
slow, critical fluctuations, of the local field near
$T_{\text{N}}$.

As $T$ decreases below 23~K, there is an increase in $T_2^{-1}$ by
the factor 2.3. From elementary considerations, one might expect a
\textit{reduction} in $T_2^{-1}$ caused by the AF field \lq\lq
detuning\rq\rq~the $^{7}$Li spins. Although it should be present,
this mechanism is clearly not the dominant one for $T_2^{-1}$.
There are three mechanisms that are often responsible for such an
increase in $T_2^{-1}$: (1) A slow fluctuation of the local
magnetic field or EFG in the frequency range near to $T_2^{-1}$,
(2) an increase in $T_1^{-1}$ to values on the order of or larger
than $T_2^{-1}$ caused by other mechanisms, or (3) an enhancement
of the spin-spin interactions between the nuclei being measured.
We expect that the first is not determining $T_2^{-1}$ because it
seems unlikely that slow fluctuations would be independent of $T$
when the fast ones that determine $T_1^{-1}$ vary so rapidly with
$T$. The second obviously does not apply because all of the
measured values of $T_1^{-1}$ are too small. It may be that the
third mechanism does apply through the $^{7}$Li-$^{7}$Li spin-spin
interaction being enhanced by the Suhl-Nakamura interaction
mediated by the AF spin wave modes\cite{hone1969} that are
expected to form at low $T$. Although this approach shows some
promise, to evaluate it in detail is beyond the scope of this
paper.

\section{Discussion}

In this Section, we discuss and interpret features of our results
that have not been covered in the above presentation of the data.

\subsection{Magnetic structure}\label{magnstruc}

In this Section we discuss how the NMR spectra of both the
powdered and the single crystal samples can be used to infer the
spatial arrangement of the moments in the AF phase of
LiVGe$_2$O$_6$. First, we consider what can be inferred from the
data in the randomly oriented polycrystalline sample. Its NMR
spectra in the AF phase can be analyzed in a similar way as in the
paramagnetic phase (see Sec.~\ref{spec}); i.e. sum the
contribution of all the V$^{3+}$ ions in a sphere of about 5nm
around the $^7$Li nuclei, but with the magnetic moments on the
V$^{3+}$ ions having an AF configuration. We used the AF structure
corresponding to a simple AF period along the chains, a
ferromagnetic order between the chains  and a value of
1.14~$\mu_{\text{B}}$ for the magnetic moments reported by Lumsden
et al.,\cite{lumsden2000} but with several different orientations
of the magnetic moments. For the moments pointing along the (100)
direction, a total field of only 0.022~T, which is much less than
the measured value of about 0.106~Tesla (see Sec.~\ref{spec}), is
obtained. However, when the moments are parallel to the (001)
direction, the result is 0.104~T, which is very close to the
measured value. Thus, the polycrystal measurements are compatible
with the proposed AF structure by Lumsden et
al.\cite{lumsden2000}, but with the magnetic moments pointing
along the crystallographic $c$-direction instead of the
$a$-direction.

\begin{figure}[htbp]
\includegraphics[width=85mm]{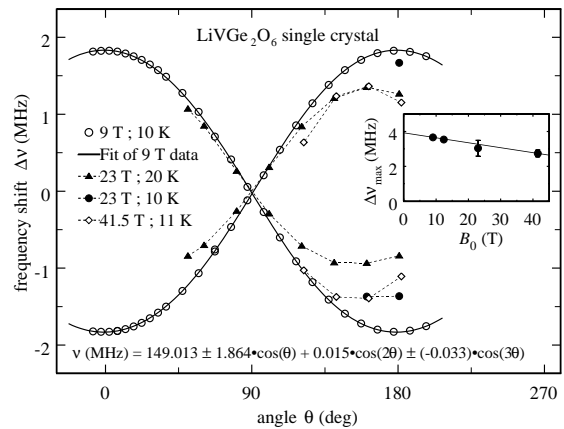}
\caption{\label{sprotb} Shift of the $^7$Li NMR-absorption peaks
in LiVGe$_2$O$_6$ as function of $\theta$ at 9.0, 23.0 and 41.5 T.
The solid lines are fits to the data at 9.0 Tesla (see text) and
the dashed lines are guides to the eye. Inset: Frequency
difference $\Delta\nu$ between the two NMR lines in the AF phase,
as function of $B_{0}$ at $T\approx 10$~K. The solid line is a
guide to the eye. }
\end{figure}

The measurements on the single crystal samples provide even more
direct evidence that the hyperfine fields at the Li sites, and
therefore the moments on the V$^{3+}$ ions, are aligned parallel
(and antiparallel) along the crystallographic $c$-direction. This
can be seen from Fig.~\ref{sprotb}, where $\Delta\nu$ for the two
peaks in the $^7$Li spectrum caused by the magnetically
inequivalent sites in the AF phase near 10~K is plotted as
function of $\theta$. For the measurements at 9.0~T, a good fit to
$\Delta\nu$ is given by:
\begin{equation}\label{rot}
  \Delta\nu=149.01\pm1.86\cos(\theta)+0.01\cos(2\theta)\mp0.03\cos(3\theta).
\end{equation}
The dominant term in this fit is proportional to $\cos(\theta)$,
which is consistent with $\mathbf{B_{\text{i}}}$ being parallel to
the $c$-axis for all values of $\theta$. Since
$\mathbf{B_{\text{i}}}$ has this orientation for the AF ordered
moments parallel to the $c$-axis, we conclude that the moments
themselves are aligned with the $c$-axis in the AF phase.

For the measurements at 9.0~T, the deviations from the cosine
function are quite small. The $\cos(2\theta)$ dependence can be
interpreted as a slight tipping of the moments by
$\mathbf{B_{0}}$.\cite{nagamiya1955} The inset of
Fig.~\ref{sprotb} shows the maximum shift $\Delta\nu_{\text{max}}$
as a function of $B_{0}$ at $T\approx 10$~K. The few high field
measurements near 10~K indicate a small reduction of the order
parameter. Because of the small number of measurements, this point
needs confirmation by a more complete set of measurements.

\subsection{Phase transition}\label{phasetrans}

As shown in Sec.~\ref{spec}, the NMR spectra of the
polycrystalline sample indicate the coexistence of the
paramagnetic and AF phases over a range of 1.5~K, centered around
$T_{\text{N}}$. It has been suggested\cite{gavilano2000} that a
first order transition is responsible for the coexistence of the
two phases. On the other hand the $T$-dependence of the splitting,
which is proportional to the order parameter, varies continuously
and smoothly to zero at the transition temperature. This behavior
indicates that the phase transition is of second order, or at
most, very weakly first order. An alternative explanation for the
coexistence of the two phases is a distribution of transition
temperatures $T_{\text{N}}$. Such a distribution, caused by
dislocations, stacking faults and V vacancies, could easily be
present in a polycrystalline sample. This explanation is also
consistent with the small value of the exponent $\beta$ near the
transition (inset of Fig.~\ref{t1503l}), which indicates a very
broad distribution of $T_1^{-1}$ about $T_{\text{N}}$, where
$T_1^{-1}$ varies more  rapidly with $T$ than at temperatures
nearby. The variation of $T_1^{-1}$ with $T$ very close to the
transition (inset, Fig.~\ref{LiT1linlin}) is slower than expected
given the divergent critical behavior near a second order
transition. As indicated earlier, this suppression of the critical
divergence could be caused by a distribution of $T_{\text{N}}$.

This interpretation in terms of a distribution of $T_{\text{N}}$
is supported by our measurements on single crystals. In comparison
with the polycrystalline sample, the data imply a coexistence of
both phases over the substantially narrower range of about 0.3~K
and a peak in $T_1^{-1}$ that is narrower and higher near
$T_{\text{N}}$ (inset, Fig.~\ref{t1or}). This behavior is
qualitatively consistent with a distribution of $T_{\text{N}}$ in
which the single crystal has fewer imperfections and therefore a
more narrow distribution of $T_{\text{N}}$ than the
polycrystalline sample.

In Sec.~\ref{magnstruc} it was shown that the AF state of
LiVGe$_2$O$_6$ has a rather simple magnetic structure. It has,
however, some unusual features which we turn to now. First,
consider the magnitude of $T_{\text{N}}$. We start with a simple
spin Hamiltonian which, as will be discussed later might not be
sufficient to describe the system:
\begin{equation}\label{hamsakai}
  H=J_{\parallel}\sum_{\langle i,j\rangle}S_i\cdot S_j+D\sum_{j}
  (S_j^z)^2+J_{\perp}\sum_{(i,j)}S_i\cdot S_j~,
\end{equation}
where $\langle i,j\rangle$ denotes an intrachain nearest-neighbor
pair and $(i,j)$ denotes an intrachain one. $J_{\parallel}$ and
$J_{\perp}$ are respectively the the intrachain and interchain
coupling constants, and $D$ is the single-ion anisotropy. The
value about 45~K has been estimated for
$J_{\parallel}$.\cite{millet1999,gavilano2000} Because of the
quasi 1-D character of the crystal structure, it is surprising
that the measured value $T_{\text{N}}\sim25$~K could be more than
half $J_{\parallel}$. On the basis of a mean-field
calculation\cite{scalapino1975} and the assumptions
$J_{\parallel}=45$~K, four neighboring chains, and
$T_{\text{N}}=25$~K, $J_{\perp}$ has been
estimated\cite{lumsden2000} to be about 1.4~K, or
$[J_{\perp}$/$J_{\parallel}]\simeq0.03$. More advanced
calculations\cite{sakai1990,senechal1993,koga2000} show, however,
that such a ratio is barely enough to induce an AF transition, and
very unlikely to have a value of $T_{\text{N}}$ as large as
$J_{\parallel}/2$. In fact based on the calculation of
S\'{e}n\'{e}chal et al.\cite{senechal1993} we estimate the ratio
of the interchain to the intrachain coupling constant which would
be needed to explain the relatively high transition temperature to
be larger than 0.06. From crystal considerations, however, this
value seems to be very large. For example, AgVP$_2$S$_6$, which is
a compound with a structure similar to LiVGe$_2$O$_6$, has a ratio
$J_{\perp}/J_{\parallel}\leq 10^{-5}$.\cite{Mutka1991} Therefore,
it seems that the relatively high transition temperature cannot be
explained by the interchain coupling alone. If, however, the
Haldane gap is not present, even a very small coupling between the
chains might be enough to induce an AF phase transition. Later, we
will consider this possibility in terms of a large single-ion
anisotropy.

Another remarkable feature of LiVGe$_2$O$_6$ is the weak influence
of $B_{0}$ on the properties of the AF state. This applies to
$T_{\text{N}}$, is seen in the properties of an apparent gap in
the fluctuations responsible for $T_1^{-1}$ (to be discussed
later), and is evident in the absence of a spin-flop transition
for $B_{0}$ as high as 44.5~T. The last point is noteworthy
because on the basis of a simple mean-field approximation, the
spin-flop field ($H_{\text{SF}}$) is expected to
be\cite{foner1963}
\begin{equation}\label{spinflop1}
  H_{\text{SF}}=(2H_{\text{E}}H_{\text{A}}-H_{\text{A}}^2)^{1/2}~,
\end{equation}
where $H_{\text{A}}$ is the anisotropy field and $H_{\text{E}}$ is
the exchange field. The field $H_{\text{A}}$ is related to the
single-ion anisotropy according to $D=H_{\text{A}}
g\mu_{\text{B}}/S$. From these considerations, $D > 50$~K, which
might be responsible for closing the Haldane gap\cite{schulz1986}
and may explain the relatively high AF-phase transition
temperature $T_{\text{N}}$. Although the mean-field approximation
used here might not be fully appropriate because $H_{\text{A}}$ is
comparable to $H_{\text{E}}$, it does allow us to obtain at least
a rough  estimate of the single-ion anisotropy using the
Hamiltonian of Eq.~\ref{hamsakai}.

It should be pointed out that according to a recent
publication,\cite{liu2001} no phase transition is expected to
occur in LiVGe$_2$O$_6$. In this work, mid-gap states are assumed
to be responsible for the susceptibility anomaly in the
experimental data of LiVGe$_2$O$_6$ and it is predicted that this
anomaly will be weaker if their are fewer crystal defects and
non-magnetic impurities. This interpretation is in clear
contradiction with the prior NMR\cite{gavilano2000} and neutron
diffraction\cite{lumsden2000} experiments and our NMR measurements
reported here. Our experiments on both polycrystalline and high
quality single crystal samples show clearly that a magnetic phase
transition occurs and permit us to refine the magnetic structure
proposed earlier.\cite{lumsden2000} Although the phase transition
appears to be more complicated than a simple ordering driven by
inter- and intrachain coupling, there is no doubt that the V
moments order antiferromagnetically below approximately 25~K.

\subsection{$T_1^{-1}$ as function of $T$,$B$, $\theta$ and
$\phi$}

Although in principle part of the coupling responsible for
$T_1^{-1}$ of the $^{7}$Li spins could be quadrupolar, the
discussion in the next paragraph argues that there is no
significant quadrupolar contribution to it. As a result, our
interpretation of $T_1^{-1}$ involves only magnetic coupling to
the $^{7}$Li spins.

There are basically two mechanisms for electric field gradient
fluctuations that might contribute to $T_1^{-1}$ of the $^{7}$Li
spins: Charge fluctuations associated with some instability of the
lattice and the Raman-type phonon process first described by van
Kranendonk.\cite{vankranendonk1954} Since the quadrupolar
splitting found in Sec.~\ref{T2} is very small and does not change
over the temperature range studied, there is no reason to expect a
significant contribution to $T_1^{-1}$ from charge fluctuations.
Also, the small $^{7}$Li quadrupolar interaction and observed
$T$-dependence of $T_1^{-1}$ rule out quadrupolar relaxation by
phonons.\cite{vankranendonk1954} We therefore conclude that there
is no significant quadrupolar contribution to $T_1^{-1}$ and that
it is caused by magnetic fluctuations.

\begin{figure}[htbp]
\includegraphics[width=85mm]{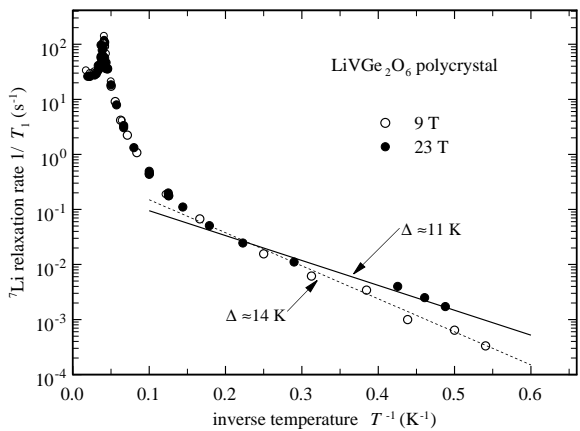}
\caption{\label{delta923} $^{7}$Li $T_1^{-1}$  as function 1/$T$
in polycrystalline LiVGe$_2$O$_6$ at 9.0 and 23~T. The solid
dashed lines are fits to the low $T$ data using Eq.~\ref{gap}. The
values shown for $\Delta$ are 14~K at 9~T and and 11~K at 23~T.}
\end{figure}

In Fig.~\ref{delta923} the spin-lattice relaxation rate,
$T_1^{-1}$, of polycrystalline LiVGe$_2$O$_6$ is plotted as
function of $T^{-1}$ for $B_{0}=9.0$~T and 23~T. Well into the AF
phase below 10~K, the behavior of $T_1^{-1}$ has the
$T$-dependence expected of electron spin excitations across an
energy gap ($\Delta$); i.e. the slope of the curve is constant and
negative at low $T$. If one simply fits the data to
\begin{equation}\label{gap}
  T_1^{-1}(T)\propto \exp(-\Delta /T)~,
\end{equation}
the values obtained for $\Delta$ are 14~K at 9.0~T and 11~K
at 23~T.

Such values must, however, be interpreted as an average over a
distribution of $\Delta$ that is quite broad. This is seen in
Fig~\ref{t1orient3}, where there is a large variation of the slope
for two different values of $\phi$ at 28.5~T (solid lines), and a
large slope at 9~T and $\phi\simeq0$ (dashed). The two values
$\Delta =$2.6~K and $\Delta =$14.7~K indicate a very large effect
of the alignment of $B_0$ in the azimuthal plane on $\Delta$ for
magnetic excitations at low $T$. Therefore, the following
discussion of the polycrystalline sample results applies to an
average behavior and is approximate and phenomenological.

First, consider the effect of $B_0$ on $\Delta$. In a 3-d
antiferromagnet, when there is a gap in the excitation spectrum,
it usually depends strongly on $B_0$ because of the Zeeman
interaction, which is $\propto g\mu_{B}B_0$. Instead, our
measurements show the very weak field dependence of less than 3~K
for a difference in applied field of 14 Tesla. Since
$g\mu_{\text{B}}B/k_{\text{B}}$, with $g=1.79$, is about 28~K for
$B = 23$~T, the usual expectations for the dependence of $\Delta$
on $B_0$ do not apply.

Also, in comparison to our results in Sec.~\ref{phasetrans},
$\Delta$ appears too small to be attributed to a magnon gap. From
$\Delta =$ 14~K and 11~K at 9.0~T and 23~T respectively, one would
expect a $\Delta\simeq16$~K. From this value, a spin-flop field
smaller than 20~T is expected, which is contradicted by the
absence of a spin-flop transition up to 44.5~T shown by our
NMR-spectrum measurements. On the basis of this evidence, it
appears that at low $T$ the magnetic fluctuations responsible for
$T_1^{-1}$ in LiVGe$_2$O$_6$ are not simple magnon excitations.
This is particularly evident from the $T_1^{-1}$ measurements
shown in Figs.~\ref{t1orient3}~and~\ref{t1orient4}, where very
large anisotropies in $T_1^{-1}$ are seen.

It is rather difficult to identify the microscopic mechanism
responsible for the very large anisotropy of $T_1^{-1}$ shown in
Fig.~\ref{t1orient4}, where, for comparison, the variation
$\cos^{4}\phi$ and $\cos^{8}\phi$ are shown by the dashed and
solid lines, respectively. For example, if it were caused by a
fluctuating magnetic field aligned with the value of $\phi$
corresponding to the maximum in $T_1^{-1}$, for other values of
$\phi$, one would expect the much weaker angular variation
$T_1^{-1}\propto \cos^{2}\phi$. Similar arguments for quadrupolar
relaxation (excluded above from other arguments) by a fluctuating
EFG could give a variation up to $\cos^{4}\phi$.

Although we do not have a microscopic model for it, the
temperature and angular variations seen in
Figs.~\ref{t1orient3}~and~\ref{t1orient4} suggest a gap-type
behavior with a splitting which itself is very anisotropic. More
work, both experimental and theoretical, is needed to identify the
mechanisms for this behavior.

In summary, the angular dependence of the spin-lattice relaxation
rate measurements for the single crystal samples is rather
complicated and it is difficult to construct a detailed
interpretation. For both the paramagnetic regime and the AF regime
down to about 8~K, $T_1^{-1}$ Fig.~\ref{t1angle} and
Eq.~\ref{sin2} indicate that the largest  contribution to
$T_1^{-1}$ has the angular dependence $\propto\sin^{2}\theta$.
This behavior is consistent with magnetic fluctuations that are
predominantly along the $c$-direction that have dipolar coupling
and only a small isotropic contribution. It may reflect primarily
amplitude fluctuations of the AF order parameter.

Below about 8~K, the disappearance of the $\theta$-dependence of
$T_1^{-1}$ and the emergence of its $\phi$-dependence indicates
that at low $T$, the origin of the fluctuations responsible for
$T_1^{-1}$ is very different from what it is at higher $T$.

\subsection{Orbital degrees of freedom and magnetic anisotropy}

In this section we mention some points about the magnetic
anisotropy and the orbital degrees of freedom of the V electrons
in LiVGe$_2$O$_6$ suggested by our measurements. They suggest that
the second order splitting $\Delta_{\text{CF}}$ of the
$t_{\text{2g}}$ orbitals is rather small, with the consequence
that (i) there is a large uniaxial anisotropy as
$g_{\perp}=2(1-\lambda/\Delta_{\text{CF}})$ ($\lambda =$
spin-orbit coupling) is strongly reduced for small values of
$\Delta_{\text{CF}}$, and (ii) orbital fluctuations may play a
significant role in the properties of $T_1^{-1}$.

A large anisotropy energy has not yet been observed directly from
$^{51}$V NMR measurements because we were not able to observe the
signal. It can, however, be inferred from several aspects of our
$^7$Li NMR measurements. In particular, the absence of a
significant $B_0$-dependence of $T_{\text{N}}$
(Fig.~\ref{TLHorderpar}), the absence of a spin-flop transition
for $B_0$ up to 44.5 T, and the $\theta$ dependence of $\Delta\nu$
(Fig.~\ref{sprotb}) all indicate a large uniaxial anisotropy for
the static magnetization. A similar picture emerges from the
$T_1^{-1}$ measurements above about 8~K, where the
$\theta$-dependence of $T_1^{-1}$ (Fig.~\ref{t1angle}) indicates
that the fluctuations of the V moments are constrained mainly to
the $c$-direction.

The situation of the V ions in LiVGe$_2$O$_6$ is similar to those
in V$_2$O$_3$, where it has been reported\cite{Takigawa1996} that,
at least for the metallic phase, $T_1^{-1}$ at the V sites is
dominated by orbital fluctuations. In LiVGe$_2$O$_6$ the presence
of orbital fluctuations of the V$^{3+}$ $t_{\text{2g}}$ orbitals
is suggested by the azimuthal $\phi$ dependence of the
spin-lattice relaxation rate at the $^7$Li site
(Figs.~\ref{t1orient3}~and~\ref{t1orient4}) which is very
anisotropic and independent of the magnitude of $B_0$.
Furthermore, orbital fluctuations might be very effective at the
$^{51}$V site itself, and may be responsible for the absence of
the $^{51}$V NMR signal.

According to a quantum chemistry analysis of
LiVGe$_2$O$_6$,\cite{millet1999,gavilano2000} if
$\Delta_{\text{CF}}$ between the V$^{3+}$ orbitals is similar to
the hopping integrals between neighboring V sites, it is not
possible to describe the system with a pure spin Hamiltonian and
the orbital degrees of freedom have to be included explicitly. Our
NMR data indicate that this is the case in LiVGe$_2$O$_6$.
Therefore, the Hamiltonian given in Eq.~\ref{hamsakai} should be
modified to better describe the physics of LiVGe$_2$O$_6$, and
include orbital degrees of freedom which might or might not be
strongly coupled to the spin degrees of freedom.

\section{Conclusions}

We have presented pulsed $^{7}$Li NMR measurements in
polycrystalline and single crystal samples of the quasi
one-dimensional antiferromagnet LiVGe$_2$O$_6$ over the $B_{0}$
and $T$ ranges 9-44.5~T and 1.5-300~K respectively. They cover
both the paramagnetic and the AF phases, for which the transition
is at $T_{\text{N}}\simeq 24.5$~K. The measurements include NMR
spectra and the relaxation times $T_1^{-1}$ and $T_2^{-1}$, often
as a function of alignment of $B_{0}$. From the spectrum
measurements, we find that in the AF phase the magnetic structure
is consistent with that reported on the basis of neutron
diffraction measurements,\cite{lumsden2000} but with the moments
aligned parallel to the $c$-axis. Measurements of $T_2^{-1}$ show
oscillations caused by the static electric field gradient. The
corresponding interaction is quite small and independent of $T$,
which indicates that over the range of $T$ that was covered, no
lattice structural transition is observed in LiVGe$_2$O$_6$. The
spectrum measurements also provide the $T$-dependence of the order
parameter and show that the transition is either second order or
weakly first order. The coexistence of the two phases over a
narrow range around $T_{\text{N}}$ and the behavior of the NMR
linewidth below it is attributed to a distribution of
$T_{\text{N}}$ in the samples. Both the spectra and the angular
dependence of $T_1^{-1}$ indicate that the external field has at
most a small effect on the alignment of the AF moment. There is no
spin-flop transition up to 44.5~T. These features show that there
is a very large anisotropy energy in this material and that the
Hamiltonian should include orbital degrees of freedom to
adequately describe it. Below 8~K, a rapid dependence of
$T_1^{-1}$ on the azimuthal angle suggests the presence of a low
energy gap for magnetic fluctuations that is highly anisotropic.

\section{Acknowledgements}
The UCLA part of the work was supported by NSF Grants DMR-0072524.
Work performed at the National High Magnetic Field Laboratory was
supported by the National Science Foundation under Cooperative
Agreement No. DMR-9527035 and the State of Florida. One of us
(A.G.) was supported by the National Institute for Materials
Science, Tsukuba, Ibaraki, 305-0003 Japan.

\end{document}